\documentclass[sigplan,10pt,nonacm]{acmart}         
\makeatletter                   
\def\mdseries@tt{m}             
\makeatother        

\makeatletter
\renewcommand\@formatdoi[1]{\ignorespaces}
\makeatother

\usepackage[export]{adjustbox}
\usepackage{listings}
    
\usepackage{color}                                                                   
\usepackage{xcolor}                                            
\usepackage{colortbl}                       
\usepackage{changepage} 
                                                                                     
\usepackage{url}                                                                     
\hypersetup{hidelinks=false,colorlinks=true,linkcolor=blue,urlcolor=blue,citecolor=blue,anchorcolor=blue} 
\usepackage{breakurl}                                                                

\usepackage{tabularx}
\usepackage{booktabs} 

\usepackage{cleveref}
\crefformat{section}{\S#2#1#3}
\crefformat{subsection}{\S#2#1#3}
\crefformat{subsubsection}{\S#2#1#3}
\crefrangeformat{section}{\S\S#3#1#4 to~#5#2#6}
\crefmultiformat{section}{\S\S#2#1#3}{ and~#2#1#3}{, #2#1#3}{ and~#2#1#3}

\usepackage{microtype}
\usepackage[utf8]{inputenc}                                                          

\usepackage[iso,american,cleanlook]{isodate}    

\usepackage{rviewport}                                                               

\usepackage{tikz} 
\usepackage[inline]{enumitem} 

\newcommand\Invisible[1]{                                                            
  \marginpar{\color{white}{\fontsize{.5}{.5}\selectfont #1 }}                        
}

\newcommand{\Exclude}[1]{}

\definecolor{Gray95}{gray}{0.95}
\definecolor{forestgreen}{rgb}{0.13, 0.55, 0.13}
                                                                                     
\usepackage[scaled]{beramono} 
\usepackage[T1]{fontenc}

\newcommand{\AtFoot}[1]{\let\thefootnote\relax\footnotetext{{#1}}}

\acmConference[]{}{}{}
\acmYear{}
\acmISBN{} 
\acmDOI{} 
\startPage{1}

\usepackage{booktabs}   
\usepackage{subcaption} 

\newcommand{\orcidicon}[1]{\href{https://orcid.org/#1}{\includegraphics[scale=0.06]{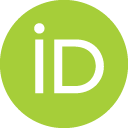}}}

\begin{document}

\title[]{Intra-process Caching and Reuse of Threads} 


\author{Dave Dice \orcidicon{0000-0001-9164-7747}}
\orcid{0000-0001-9164-7747}             
\affiliation{
  \institution{Oracle Labs}             
}

\email{first.last@oracle.com}            

\author{Alex Kogan \orcidicon{0000-0002-4419-4340}} 
\orcid{0000-0002-4419-4340} 
\affiliation{
  \institution{Oracle Labs}             
}
\email{first.last@oracle.com}          



\begin{abstract}

Creating and destroying threads on modern Linux systems incurs high latency, absent concurrency, and
fails to scale as we increase concurrency.   
To address this concern we introduce a process-local cache of idle threads.  Specifically, instead of destroying
a thread when it terminates, we cache and then recycle that thread in the context of subsequent thread creation requests.
This approach shows significant promise in various applications and benchmarks that create and destroy 
threads rapidly and illustrates the need for and potential benefits of improved concurrency infrastructure.  
With caching, the cost of creating a new thread drops by almost an order of magnitude. As our experiments demonstrate, 
this results in significant performance improvements for multiple applications that aggressively create and destroy numerous threads.

\end{abstract}

\ccsdesc[300]{Software and its engineering~Multithreading}
\ccsdesc[300]{Software and its engineering~Concurrency control}


\keywords{Threads, Concurrency Control}  

\maketitle

\thispagestyle{fancy}

\section{Introduction}

To mitigate the costs of thread creation and destruction, we implement a user-mode process-local
cache of \emph{idle} threads.  These are threads that have logically terminated, but which 
we capture and retain for subsequent reuse.  
Threads that would normally exit to the kernel are now retained and reused, sparing the cost of 
creating new threads in the future
Using various benchmarks, we show that our approach can yield improved performance. 

\section{Implementation} 

We implemented our proof-of-concept cache as an \texttt{LD\_PRELOAD} interposition module which intercepts POSIX 
\cite{POSIX,PThreads}
\texttt{pthread\_\allowbreak{}create, pthread\_\allowbreak{}exit, pthread\_\allowbreak{}detach and pthread\_\allowbreak{}join} 
calls, reimplementing those operators as necessary
to provide a thread cache.  Using \texttt{LD\_PRELOAD} interposition allows us to enable or disable the cache by
simply setting an environment variable, and allows us to use unmodified application executables. 
Logically terminated threads are retained on a local \emph{Idle} list.  
When creating new threads, our implementation first attempts to allocate a thread from the idle list.  
Failing that, it falls back to the underlying \texttt{pthread\_\allowbreak{}create} API.  
The idle list is maintained in a LIFO fashion to better leverage residual cache residency and scheduling affinity.

A \emph{physical} thread is assigned various thread-private resources, including user-mode structures, a user-mode stack, 
kernel-mode structures, and a kernel-mode stack.  Creating and destroying a physical thread entails creating and destroying
those constituent parts.  By retaining idle physical threads in our cache, we can avoid those costs.  As a consequence
the cache also significantly reduces time spent executing in kernel mode.

\section{Performance Evaluation} 

All experiments were run on an Oracle X5-2.  The system has 2 sockets, each populated with 
an Intel Xeon E5-2699 v3 CPU running at 2.30GHz.  Each socket has 18 cores, and each core is 2-way 
hyperthreaded, yielding 72 logical CPUs in total.  The system was running Ubuntu 20.04 with a stock 
Linux version 5.4 kernel, and all software was compiled using the provided GCC version 9.3 toolchain
at optimization level ``-O3''.  
64-bit C or C++ code was used for all experiments.  
Factory-provided system defaults were used in all cases, and Turbo mode was left enabled.  
In all cases default free-range unbound threads were used.  

In Table-\ref{Table:Performance} we compare the performance of the default system against the
same system with our cache activated.  The \texttt{Spawn} benchmark creates 32 concurrent threads.
We use C++ \texttt{std::thread} where the GNU/Linux implementation maps each such thread 1:1 to
an underling native POSIX \texttt{pthread}.  
Each of those threads loops, creating an additional thread and then waiting for that thread to join. 
The threads created in the loop exit immediately.  
Each of the 32 ``creator'' threads is independent and there is no communication or synchronziation within this set of threads. 
At the end of a 10 second measurement interval the benchmark reports the number of threads spawned per second.
We report the median of 7 runs.  (We use the median of 7 independent runs for all results shown 
in this table).  The \texttt{stdasync} benchmark is similar but uses the newer C++ \texttt{std::async} 
construct instead of explicit threads.  GNU/Linux implements each \texttt{std::async} instance
as a native POSIX thread.  

\texttt{HuggingFace} is a machine learning BERT language model \cite{BERT} 
inference benchmark provided with the huggingface 
transformer models package \footnote{\url{https://huggingface.co/transformers/benchmarks.html}} 
running on \texttt{Pytorch} version 1.5 \footnote{The command line was as follows: 
python3 examples/benchmarks.py --torch --models bert-base-cased  --no\_memory --batch\_sizes 1  --slice\_sizes 64}.  As used by PyTorch, the underlying GNU OpenMP (libgomp) spawns and terminates threads rapidly, 
impacting performance of the inference benchmark. 

All the remaining benchmarks are from the Inncabs\cite{inncabs} benchmark
suite, which is designed to measure the performance of the \texttt{std::async} construct. 
We obtained the source code from \url{https://github.com/PeterTh/inncabs}
\footnote{A number of the Inncabs benchmarks failed in default mode and were thus excluded.}. 
The implementation of the C++ \texttt{std::async} construct defers creating an 
underlying thread until the returned \texttt{promise} is evaluated.  
All Inncabs benchmarks report elapsed time in milliseconds. 

\begin{figure}[h!]
\includegraphics[width=8.5cm]{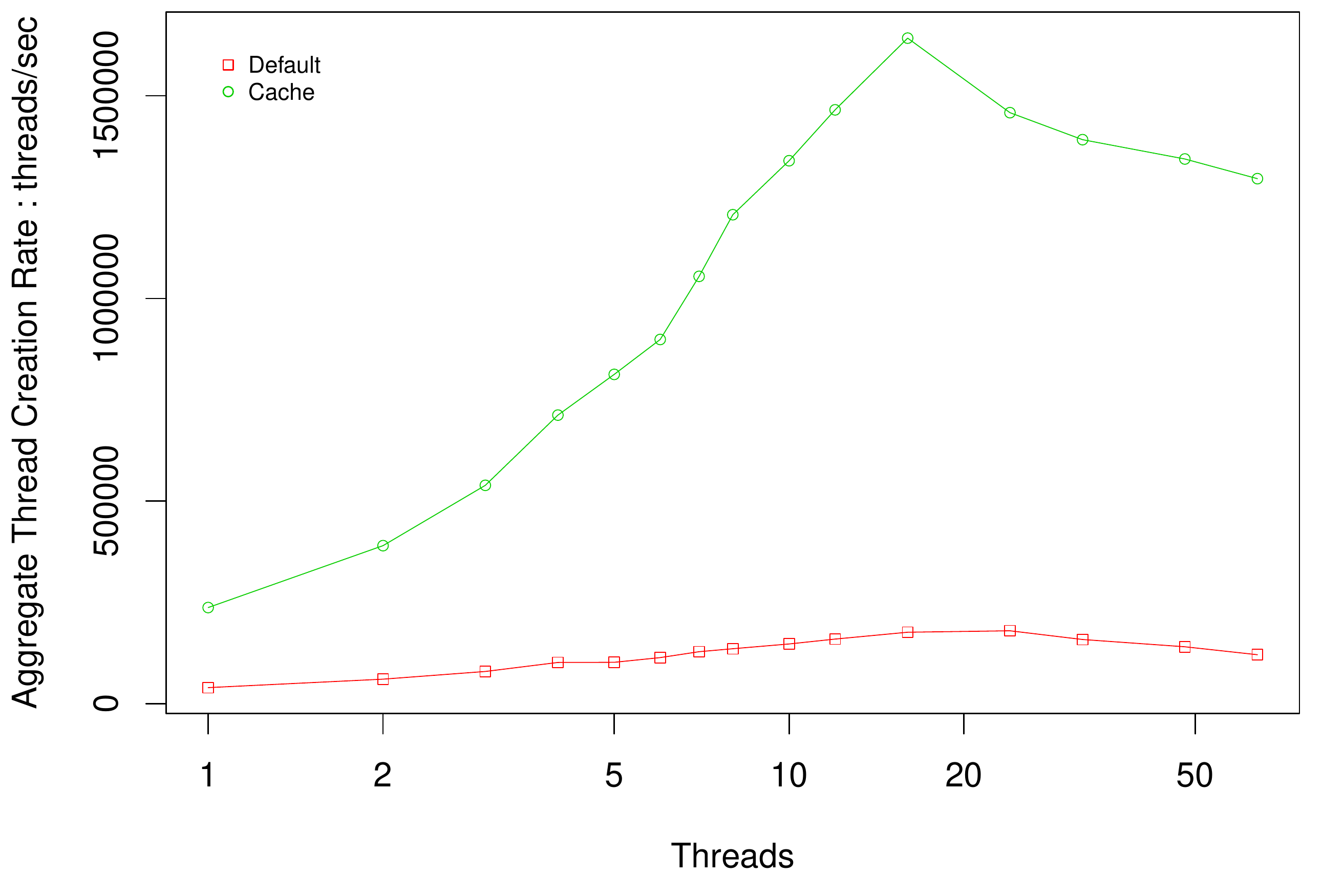}
\vspace{-18pt}      
\caption{Thread Creation Scalability}
\label{Figure:scale}
\end{figure}

In Figure-\ref{Figure:scale} we compare the scalability of thread creation
using the \texttt{spawn} benchmark, above, varying the number of threads on the X-axis
and showing aggregate thread creation rates on the Y-axis.  
For clarity and to convey the maximum amount of information to allow a comparision the algorithms,
the $Y$-axis is logarithmic.
\texttt{Default} reflects default thread creation while \texttt{cache} reflects performance when
the thread cache is enabled.  As we can see, the cache decreases latency and also improves
scalability.  At one thread, the benchmark serves as simple measure of unloaded latency for
creating and destroying threads.  The cache improves performance by 5.9 times at 1 thread. 

\Invisible{concepts: physical thread vs logical thread} 


\Invisible{So instead of creating a new \emph{physical} thread, the cost with the cache is instead just waking up an existing 
idle thread in order to dispatch new \emph{logical} work assignment.  Context switching, while bad, is still faster and 
more scalable than thread creation. } 

\Invisible{
I took some data yesterday and today on a variation that doesn’t reuse/recycle threads, 
but rather tries to keep a pool of ready-to-launch nascent standby threads.   
When we intercept a pthread\_create() call we try to first extract an element from the pool 
and dispatch the work there.   Failing that, we launch the classic hard way by creating a new thread
right then and there.  When thread exits, we spawn a new replacement thread that enters itself 
into the pool to back-fill the element we removed.   We’re still creating and destroying 
kernel threads frequently, but this approach attempts to get some of the overheads out 
of the critical paths, and allow us to launch faster.   The advantage to this approach 
is that I’d argue it’s completely safe and sound as we never reuse physical threads.   
It’s also pretty simple — or at least simpler than the accelerator where we reuse threads.

The performance results are muddy.   If the threads are very short-lived  then we just 
can’t keep up as the kernel just can’t produce and destroy threads at a rate sufficient to 
keep up with our demand to back-fill the pool.   I think we’re just rate-limited by 
kernel bottlenecks.   In this case there’s very little benefit behind the default baseline.    
If the threads are a more longer-lived then there’s some measurable benefit and performance 
is better than the baseline default, but still worse by a large factor than the implementation 
where we reuse threads.  

I tried numerous variations with parallel/helping back-filling, etc, to try to hide or 
amortize thread creation latency but came to the conclusion we’re just limited by the kernel.

I looked quickly at the stack caching approach, and some of the thread/stack APIs we used 
in the past are now deprecated as unreliable.  It’s still possible to make it work, but we 
need to manage thread stacks fully end-to-end.   
}

\Invisible{Capture and retain moribund threads and reuse them in response to subsequent 
thread creation requests, thus avoiding kernel overheads.  
}

\Invisible{
Invention disclosure accession number :
Oracle File Number: IDF-133336 
Title of Invention: Intra-process Caching and Reuse of Threads 
ORC2133336-US-NPR (Intra-process Caching and Reuse of Threads)
}


\begin{table} [h]
\centering
{\fontsize{6.5}{6.5}\selectfont
\begin{tabular}{lrrr}
\toprule

\multicolumn{1}{r}{Default}      &
\multicolumn{1}{r}{With Cache}   &
\multicolumn{1}{r}{Units}        \\
\midrule

Spawn           & 158940  & 1399817  & threads/second  \\ 
stdasync        & 132993  & 1354640  & threads/second  \\
HuggingFace     & 94      & 41       & milliseconds    \\ 
Alignment 100   & 872     & 314      & milliseconds    \\
SparseLRU       & 620     & 362      & milliseconds    \\
Sort 1000000    & 770     & 41       & milliseconds    \\
Health small    & 60669   & 2713     & milliseconds    \\
Floorplan       & 3431    & 974      & milliseconds    \\
FFT 100000      & 3860    & 484      & milliseconds    \\

\midrule[\heavyrulewidth]
\bottomrule
\end{tabular}%
}
\caption{Performance Comparison}\label{Table:Performance}
\end{table}

\section{Future Directions: Cache Retention Policies}
Our preferred implementation simply caches all terminated threads.  If the application
has a maximum of $N$ concurrently live threads, then the worst case retention in the idle cache is $N-1$ threads.
Other more refined policies are possible, such as (a) clamping the number of threads in the cache; (b) 
aging out and culling threads idle threads that have not run in some tunable period, or (c) bounding the
thread $\cdot$ seconds integral of the list and culling until the target is limit is reached.  
To reduce the memory footprint of cached threads, the implementation might also use the \texttt{madvise(MAP\_DONTNEED)} 
advisory service to inform the operating system that the dirty pages on an idle stack may be reclaimed and replaced
lazily by demand-zero-fill pages.  

All the policies above are process-local, but more sophisticated system-wide polices may be more effective, 
where processes cooperate, possibly informed by a measure of system memory pressure, to decide to trim their caches.





\bibliography{ThreadCache.bib}


\begin{thebibliography}{4}


\ifx \showCODEN    \undefined \def \showCODEN     #1{\unskip}     \fi
\ifx \showDOI      \undefined \def \showDOI       #1{#1}\fi
\ifx \showISBNx    \undefined \def \showISBNx     #1{\unskip}     \fi
\ifx \showISBNxiii \undefined \def \showISBNxiii  #1{\unskip}     \fi
\ifx \showISSN     \undefined \def \showISSN      #1{\unskip}     \fi
\ifx \showLCCN     \undefined \def \showLCCN      #1{\unskip}     \fi
\ifx \shownote     \undefined \def \shownote      #1{#1}          \fi
\ifx \showarticletitle \undefined \def \showarticletitle #1{#1}   \fi
\ifx \showURL      \undefined \def \showURL       {\relax}        \fi
\providecommand\bibfield[2]{#2}
\providecommand\bibinfo[2]{#2}
\providecommand\natexlab[1]{#1}
\providecommand\showeprint[2][]{arXiv:#2}

\bibitem[\protect\citeauthoryear{Butenhof}{Butenhof}{1997}]%
        {PThreads}
\bibfield{author}{\bibinfo{person}{David~R. Butenhof}.}
  \bibinfo{year}{1997}\natexlab{}.
\newblock \bibinfo{booktitle}{\emph{Programming with POSIX Threads}}.
\newblock \bibinfo{publisher}{Addison-Wesley Longman Publishing Co., Inc.}
\newblock
\showISBNx{0201633922}


\bibitem[\protect\citeauthoryear{Devlin, Chang, Lee, and Toutanova}{Devlin
  et~al\mbox{.}}{2018}]%
        {BERT}
\bibfield{author}{\bibinfo{person}{Jacob Devlin}, \bibinfo{person}{Ming{-}Wei
  Chang}, \bibinfo{person}{Kenton Lee}, {and} \bibinfo{person}{Kristina
  Toutanova}.} \bibinfo{year}{2018}\natexlab{}.
\newblock \showarticletitle{{BERT:} Pre-training of Deep Bidirectional
  Transformers for Language Understanding}.
\newblock \bibinfo{journal}{\emph{CoRR}}  \bibinfo{volume}{abs/1810.04805}
  (\bibinfo{year}{2018}).
\newblock
\showeprint[arxiv]{1810.04805}
\urldef\tempurl%
\url{http://arxiv.org/abs/1810.04805}
\showURL{%
\tempurl}


\bibitem[\protect\citeauthoryear{IEEE and Group}{IEEE and Group}{[n. d.]}]%
        {POSIX}
\bibfield{author}{\bibinfo{person}{IEEE} {and} \bibinfo{person}{The~Open
  Group}.} \bibinfo{year}{[n. d.]}\natexlab{}.
\newblock \bibinfo{title}{The Open Group Base Specifications Issue 7, 2018
  edition; IEEE Std 1003.1-2017 (Revision of IEEE Std 1003.1-2008)}.
\newblock
\newblock


\bibitem[\protect\citeauthoryear{{Thoman}, {Gschwandtner}, and
  {Fahringer}}{{Thoman} et~al\mbox{.}}{2015}]%
        {inncabs}
\bibfield{author}{\bibinfo{person}{P. {Thoman}}, \bibinfo{person}{P.
  {Gschwandtner}}, {and} \bibinfo{person}{T. {Fahringer}}.}
  \bibinfo{year}{2015}\natexlab{}.
\newblock \showarticletitle{On the Quality of Implementation of the C++11
  Thread Support Library}. In \bibinfo{booktitle}{\emph{2015 23rd Euromicro
  International Conference on Parallel, Distributed, and Network-Based
  Processing}}.
\newblock
\urldef\tempurl%
\url{https://doi.org/10.1109/PDP.2015.33}
\showDOI{\tempurl}


\end{thebibliography}




\section{Appendix : Additional Remarks}

Various strategies to accelerate thread creation are possible.  As noted above
an implementation might cache threads proper.  Caching of thread stacks is also a viable approach
and has been used in the Solaris Operating Environment.  An implementation might also avoid
recycling threads, but instead keep an anticipatory cache of ready-to-run \emph{standby} threads available for
immediate dispatch.

Caching threads via an \texttt{LD\_PRELOAD} module is not strictly sound, as, for instance,
thread-local storage elements are not reset when a physical thread recycles.   While we have not
observed any errors related to this concern, we note that an implementation that caches
threads is best implemented in the threading library proper which has access to the thread
local storage elements and is able to reset those values in the expected fashion.
Recycling threads also allows \texttt{pthread\_self} thread identifiers to recycle rapidly,
possibly causing latent application errors (related to holding potentially stale thread identifiers)
to manifest more frequently.  

We currently implement the idle list is a stack protected by a lock.  
Using per-NUMA node idle lists appears to be a valid implementation optimization.  
In addition, since the list can be subject to high traffic, a lock-free stack may be more appropriate, or a
"half lock-free" approach where \texttt{push} operations are implemented via an atomic compare-and-swap (CAS) loop
and \texttt{pop} are also implemented via atomics but protected by a lock to avoid the \emph{A-B-A} pathology.

\end{document}